\documentclass{article}
\usepackage{graphicx}
\usepackage{amsmath}
\usepackage{amssymb}
\usepackage{authblk}
\usepackage{tikz}
\usepackage{float}
\usepackage{booktabs}
\usepackage{url}
\usetikzlibrary{shapes, positioning, fit}
\usepackage[margin=0.5in]{geometry}

\author[1]{Mark M. Bailey \thanks{corresponding author: mark.m.bailey@ni-u.edu}}
\author[2]{Matthew J. Hasenjager}
\author[3]{Nina H. Fefferman}
\affil[1]{Department of Cyber Intelligence and Data Science, National Intelligence University, Bethesda, MD, USA}
\affil[1]{Biological and Computational Intelligence Center, National Intelligence University, Bethesda, MD, USA}
\affil[2]{Department of Ecology and Evolutionary Biology, University of Tennessee, Knoxville, TN, USA}
\affil[2]{National Institute for Modeling Biological Systems, University of Tennessee, Knoxville, TN, USA}
\affil[3]{School of Mathematical and Statistical Sciences, Arizona State University, Phoenix, AZ, USA}
\affil[3]{NSF Center for Analysis and Prediction of Pandemic Expansion (APPEX), Arizona State University, Phoenix, AZ, USA}

{
    \makeatletter
    \renewcommand\AB@affilsepx{: \protect\Affilfont}
    \makeatother

    \affil[ ]

    \makeatletter
    \renewcommand\AB@affilsepx{, \protect\Affilfont}
    \makeatother
}

\title{Targeted Disruption of Hypernetworks via Spectral Partitioning}

\date{April 2026}

\begin{document}

\maketitle

\begin{abstract}
We study hyperedge-removal strategies for suppressing contagion on synthetic hypergraphs. Hypergraphs are generated from Erdős--Rényi, Barabási--Albert, and Watts--Strogatz seed graphs by promoting maximal cliques to hyperedges. For each hypergraph, we construct \(s\)-line graphs whose vertices correspond to hyperedges and whose edges encode hyperedge overlap of size at least \(s\). Spectral \(k\)-way clustering of these \(s\)-line graphs yields a multiscale cut-persistence score used to rank hyperedges for removal.

Simulations show that the effect of this intervention is strongly topology-dependent. In the reported Erdős--Rényi case, cut-persistence targeting reduces final infection size more than random hyperedge removal. In the Watts--Strogatz and Barabási--Albert cases, however, random removal is comparable to or better than cut-persistence targeting. These results suggest that spectral overlap structure can identify structurally salient hyperedges, but structural salience alone does not guarantee optimal contagion suppression. The study motivates further comparison with ensemble-level experiments and explicitly higher-order contagion models.
\end{abstract}



\section{Introduction}

The complex set of interactions characterizing modern social systems presents both opportunities and challenges for understanding and influencing collective behavior. Consequently, computational social science has emerged as a pivotal field that leverages computational methods to analyze and model complex social phenomena \cite{lazer2009life}. Within this domain, network science provides a foundational framework for representing and examining the structural properties of social systems, where individuals or entities can be depicted as nodes, and their interactions as edges \cite{barabasi2016network}.

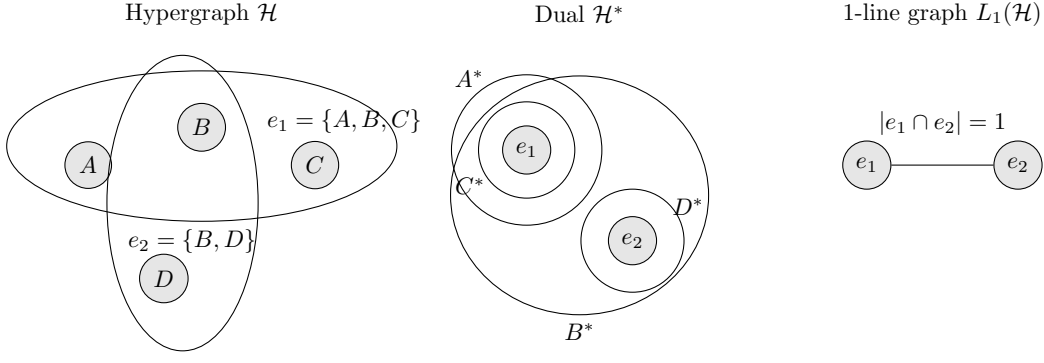
\begin{figure}[ht]
    \centering
    \begin{tikzpicture}[scale=1, every node/.style={scale=0.9}]

        \node[circle, draw, fill=gray!20] (A) at (0, 1) {$A$};
        \node[circle, draw, fill=gray!20] (B) at (1.5, 1.5) {$B$};
        \node[circle, draw, fill=gray!20] (C) at (3, 1) {$C$};
        \node[circle, draw, fill=gray!20] (D) at (1, -0.5) {$D$};

        \node[draw, ellipse, fit=(A)(B)(C), inner sep=6pt,
              label={[xshift=-4pt, yshift=-80pt]above:{$e_2=\{B,D\}$}}] {};
        \node[draw, ellipse, fit=(B)(D), inner sep=6pt,
              label={[yshift=12mm]right:{$e_1=\{A,B,C\}$}}] {};

        \node at (1.5, 3.0) {Hypergraph \(\mathcal{H}\)};

        \node[circle, draw, fill=gray!20] (Eone) at (5.8, 1.2) {$e_1$};
        \node[circle, draw, fill=gray!20] (Etwo) at (7.2, 0.0) {$e_2$};

        \node[draw, ellipse, fit=(Eone), inner sep=5pt,
              label={[yshift=3mm]above left:{$A^\ast$}}] {};
        \node[draw, ellipse, fit=(Eone), inner sep=13pt,
              label={[xshift=3mm,yshift=5mm]below left:{$C^\ast$}}] {};
        \node[draw, ellipse, fit=(Eone)(Etwo), inner sep=9pt,
              label=below:{$B^\ast$}] {};
        \node[draw, ellipse, fit=(Etwo), inner sep=6pt,
              label={[xshift=-3mm,yshift=5mm]right:{$D^\ast$}}] {};

        \node at (6.5, 3.0) {Dual \(\mathcal{H}^{\ast}\)};

        \node[circle, draw, fill=gray!20] (Lone) at (10.3, 1.0) {$e_1$};
        \node[circle, draw, fill=gray!20] (Ltwo) at (12.3, 1.0) {$e_2$};

        \draw (Lone) -- node[above,yshift=9pt] {\(|e_1\cap e_2|=1\)} (Ltwo);

        \node at (11.3, 3.0) {\(1\)-line graph \(L_1(\mathcal{H})\)};

    \end{tikzpicture}
    \caption{A hypergraph \(\mathcal{H}\), its dual \(\mathcal{H}^{\ast}\), and the \(1\)-line graph \(L_1(\mathcal{H})\). In the \(s\)-line graph used in this paper, vertices correspond to hyperedges of \(\mathcal{H}\). Two vertices are adjacent when the corresponding hyperedges overlap in at least \(s\) original vertices. Here, \(e_1\cap e_2=\{B\}\), so \(e_1\) and \(e_2\) are adjacent in \(L_1(\mathcal{H})\).}
    \label{fig1:graphs}
\end{figure}

Traditional network analyses often focus on pairwise (dyadic) interactions; however, many real-world social processes involve group interactions that transcend dyadic relationships. Hypergraphs (Figure \ref{fig1:graphs}) offer a natural extension by allowing edges (hyperedges) to connect multiple nodes simultaneously, thus capturing higher-order interactions inherent in social contexts such as group discussions, collaborative projects, and organizational structures \cite{battiston2020networks}.

Understanding the resilience and vulnerability of such complex networks is crucial, especially in scenarios where targeted interventions are necessary to disrupt harmful processes, such as the spread of misinformation or transmissible disease. Spectral clustering, particularly the $k$-way partitioning approach, has proven effective in identifying community structures within traditional networks by utilizing the eigenstructure of graph Laplacians to partition the network into cohesive subgroups \cite{von2007tutorial}. Additionally, contemporary research on network dismantling has largely emphasized edge importance in pairwise graphs. Notable contributions include Girvan and Newman's edge-betweenness method for community detection \cite{girvan2002community}, Yu et al.'s BCC ranking for identifying critical edges \cite{yu2018identifying}, and module-based attacks for graph fragmentation \cite{requiao2015fragmentation}. These techniques, however, remain difficult to transfer directly to hypergraphs without loss of higher-order structure.

Prior studies have demonstrated the utility of spectral and topological tools for network partitioning and intervention. For instance, Newman \cite{newman2006modularity} developed modularity-based methods to identify communities in large-scale networks, while Fortunato \cite{fortunato2016community} reviewed limitations and advances in community detection. Benson et al. \cite{benson2016higher} emphasized the importance of motif-based clustering in higher-order networks. Estrada and Rodriguez-Velazquez \cite{estrada2005subgraph} extended spectral graph theory to assess subgraph centrality. Traag et al. \cite{traag2011narrow} analyzed the resolution-limit problem and the restricted scope of resolution-limit-free community detection methods. More recently, Chodrow \cite{chodrow2021configuration} introduced a configuration model for higher-order systems, and Carletti et al. \cite{carletti2020random} studied random walks on hypergraphs.

Other contributions in hypergraph theory and spectral methods have extended these insights. Notably, Liu et al. \cite{liu2022high} formalized the concept of \(s\)-line graphs for non-uniform hypergraphs, providing a representation for analyzing overlaps among hyperedges based on shared node subsets. In the convention used here, the \(s\)-line graph \(L_s(\mathcal{H})\) has one vertex for each hyperedge of \(\mathcal{H}\), and two such vertices are adjacent when the corresponding hyperedges overlap in at least \(s\) original vertices. This construction allows classical graph-spectral methods to be applied to hyperedge-overlap structure without replacing hyperedges by purely dyadic interactions.

In addition, recent studies have developed scalable spectral coarsening techniques for hypergraph partitioning, further demonstrating the growing feasibility of applying spectral $k$-way clustering directly to hypergraph representations \cite{jia2024shypar}. These techniques build on earlier foundational work by Zhou et al. \cite{zhou2007learning}, who developed spectral clustering, classification, and embedding methods for hypergraphs, and Chan et al. \cite{chan2018spectral}, who studied spectral properties of hypergraph Laplacians and related approximation guarantees.

Recent contributions also underscore the importance of higher-order interactions in social transmission systems. For instance, Hasenjager an colleagues \cite{hasenjager2024social,hasenjager2026} modeled social ageing and higher-order interactions in knowledge transmission, illustrating how group-structured interactions can affect the movement of information through social systems. Studies like this emphasize the need for flexible modeling tools (like hypergraphs) that account for more than just pairwise interactions when evaluating network vulnerability and designing interventions.

In this study, we propose a methodological framework that integrates hypergraph modeling with spectral \(k\)-way partitioning to identify and target structurally salient hyperedges. Our approach involves:

\begin{enumerate}
    \item Generating hypergraphs through clique expansion of seed graphs modeled by Barabási--Albert, Erdős--Rényi, and Watts--Strogatz algorithms.
    \item Simulating contagion processes on the full hypergraph to establish baseline dynamics.
    \item Constructing \(s\)-line graphs \(L_s(\mathcal{H})\), whose vertices correspond to hyperedges of the original hypergraph.
    \item Applying spectral \(k\)-way partitioning to these \(s\)-line graphs to identify boundary hyperedges between spectral communities.
    \item Ranking hyperedges by a multiscale cut-persistence score.
    \item Removing high-ranking hyperedges and re-evaluating contagion dynamics relative to random-removal baselines.
\end{enumerate}

By analyzing the effects of hyperedge removal on contagion processes, this work provides a proof-of-concept framework for studying structural interventions in synthetic hypergraphs. The results highlight both the promise and the limitations of overlap-based spectral targeting, especially when compared with random-removal baselines.

The remainder of this paper is structured as follows. Section 2 details the theoretical background. Section 3 describes the methods, including hypergraph generation, contagion simulation, and spectral partitioning. Section 4 presents the intervention results and bifurcation scans. Section 5 discusses the implications and limitations of the findings. Section 6 concludes with directions for future work.

\section{Theoretical Background: Spectral \textit{k}-way Partitioning}

Spectral methods provide a powerful approach to graph partitioning by leveraging the eigenstructure of graph Laplacians. In our study, we apply spectral \(k\)-way partitioning to \(s\)-line graphs \(L_s(\mathcal{H})\), whose vertices represent hyperedges of the original hypergraph and whose edges encode overlap among those hyperedges.

\subsection{Spectral Partitioning via the Normalized Laplacian}

Let \( G = (V, E) \) be an undirected graph with adjacency matrix \( A \), degree matrix \( D \), and Laplacian defined as
\[
L = D - A.
\]
The \textit{normalized Laplacian} is given by
\[
\mathcal{L} = D^{-1/2} L D^{-1/2} = I - D^{-1/2} A D^{-1/2}.
\]
This matrix is symmetric and positive semi-definite, with real non-negative eigenvalues \( 0 = \lambda_1 \leq \lambda_2 \leq \cdots \leq \lambda_n \). The multiplicity of the eigenvalue \( 0 \) corresponds to the number of connected components in \( G \), and the second smallest eigenvalue \( \lambda_2 \) (the Fiedler value) reflects the graph's algebraic connectivity.

Spectral clustering leverages the eigensystem of \( \mathcal{L} \) to partition the graph into \( k \) communities. This approach approximates the solution to the \textit{normalized cut} problem, which aims to minimize the quantity
\[
\widetilde{cut}(V_1, \dots, V_k) = \sum_{i=1}^k \frac{\text{cut}(V_i, \bar{V}_i)}{\text{vol}(V_i)},
\]
where \( \bar{V}_i = V \setminus V_i \) denotes the complement of the node set \( V_i \), 
\( \text{cut}(V_i, \bar{V}_i) = \sum_{u \in V_i, v \in \bar{V}_i} A_{uv} \) is the total weight of edges leaving the partition \( V_i \), 
and \( \text{vol}(V_i) = \sum_{u \in V_i} D_{uu} \) is the volume of the partition. This formulation encourages partitions that are both sparsely connected to the rest of the graph and internally cohesive.

Following the relaxation introduced by \cite{ng2002spectral}, we compute a spectral embedding from the normalized Laplacian \(\mathcal{L}\). For a graph with \(n\) vertices, let
\[
X \in \mathbb{R}^{n\times k}
\]
denote the matrix whose columns are eigenvectors associated with the \(k\) smallest eigenvalues of \(\mathcal{L}\). The rows of \(X\) provide an embedding of graph vertices in \(\mathbb{R}^k\), and \(k\)-means clustering is then applied to these embedded points to obtain a discrete partition.

The normalized-cut problem is discrete and generally difficult to solve exactly. Spectral clustering replaces the discrete cluster-indicator constraint by a continuous orthogonality constraint. Specifically, the relaxed problem can be written as
\[
\min_{X\in \mathbb{R}^{n\times k}} 
\operatorname{Tr}(X^\top \mathcal{L}X)
\quad
\text{subject to}
\quad
X^\top X = I_k .
\]
By the Rayleigh--Ritz theorem, a minimizer is obtained by taking the columns of \(X\) to be eigenvectors associated with the \(k\) smallest eigenvalues of \(\mathcal{L}\).

In the computational implementation used for the experiments here, we compute the \(k+1\) smallest eigenvectors of the normalized Laplacian and discard the first, trivial eigenvector before applying row normalization and \(k\)-means clustering. Thus, the embedding uses the nontrivial eigenvectors associated with \(\lambda_2,\dots,\lambda_{k+1}\).

\paragraph{Relaxation bound.}
Let \(\mathcal{X}\) denote the feasible set of normalized discrete cluster-indicator matrices, and let \(X_\ast\) be the optimizer of the continuous relaxation above. Since the continuous feasible set contains the normalized discrete indicator matrices, the relaxed objective value is a lower bound on the discrete normalized-cut optimum:
\[
\operatorname{Tr}(X_\ast^\top \mathcal{L}X_\ast)
\leq
\min_{Y\in \mathcal{X}}
\operatorname{Tr}(Y^\top \mathcal{L}Y).
\]
The subsequent \(k\)-means step is a discretization heuristic; it is not guaranteed to recover the globally optimal normalized cut. This distinction is important when interpreting the resulting hyperedge partitions.

In this study, the procedure is applied to \(s\)-line graphs \(L_s(\mathcal{H})\), whose vertices are hyperedges of the original hypergraph \(\mathcal{H}\). The resulting partitions are therefore partitions of hyperedges, not partitions of original vertices.

\subsection{Application to Hypergraph-Derived \(s\)-line Graphs}

Let \(\mathcal{H}=(V,E)\) be a hypergraph with hyperedge set
\[
E=\{e_1,e_2,\dots,e_m\}.
\]
For a fixed integer \(s\geq 1\), we define the weighted \(s\)-line graph of \(\mathcal{H}\) as
\[
L_s(\mathcal{H})=(E,F_s,w_s),
\]
where the vertex set of \(L_s(\mathcal{H})\) is the hyperedge set \(E\) of the original hypergraph. Two vertices \(e_i,e_j\in E\) are adjacent in \(L_s(\mathcal{H})\) when the corresponding hyperedges overlap in at least \(s\) original vertices:
\[
\{e_i,e_j\}\in F_s
\quad\Longleftrightarrow\quad
|e_i\cap e_j|\geq s.
\]
The edge weight records the size of the overlap:
\[
w_s(e_i,e_j)=|e_i\cap e_j|.
\]
Spectral partitioning of \(L_s(\mathcal{H})\) therefore yields groupings of hyperedges according to their weighted overlap structure. In this paper, interventions are performed on hyperedges of \(\mathcal{H}\), equivalently on vertices of \(L_s(\mathcal{H})\).

\subsection{Boundary Hyperedges and Cut Persistence}

Once \(L_s(\mathcal{H})\) is partitioned into \(k\) clusters, we identify hyperedges whose corresponding vertices in \(L_s(\mathcal{H})\) lie on boundaries between spectral clusters. We call such objects \emph{boundary hyperedges}. This terminology avoids ambiguity: the cut edges of \(L_s(\mathcal{H})\) are graph edges between hyperedge-vertices, whereas the intervention objects in the original hypergraph are hyperedges.

\subsubsection{Defining Cut Persistence Score}

Let
\[
\mathcal{P}^{(s)}
=
\{C_1^{(s)},C_2^{(s)},\dots,C_k^{(s)}\}
\]
denote a spectral \(k\)-way partition of \(L_s(\mathcal{H})\), where each cluster \(C_i^{(s)}\subseteq E\) is a set of hyperedges. Let \(c_s(e)\) denote the cluster label assigned to hyperedge \(e\in E\) at overlap scale \(s\).

We say that \(e\in E\) is a boundary hyperedge at scale \(s\) if there exists another hyperedge \(f\in E\) such that \(e\) and \(f\) are adjacent in \(L_s(\mathcal{H})\) and assigned to different spectral clusters:
\[
|e\cap f|\geq s
\quad\text{and}\quad
c_s(e)\neq c_s(f).
\]
Equivalently, define the boundary indicator
\[
b_s(e)
=
\begin{cases}
1, & \text{if } \exists f\in E \text{ such that } |e\cap f|\geq s
     \text{ and } c_s(e)\neq c_s(f),\\
0, & \text{otherwise}.
\end{cases}
\]

For a set of overlap scales \(\mathcal{S}_{\mathrm{ov}}\), the cut-persistence score of \(e\) is
\[
\mathrm{CP}(e)
=
\sum_{s\in\mathcal{S}_{\mathrm{ov}}} b_s(e).
\]
Thus, \(\mathrm{CP}(e)\) counts the number of overlap scales at which \(e\) lies on a boundary between spectral communities.

Because large hyperedges may be more likely to overlap other hyperedges at multiple values of \(s\), a normalized cut-persistence score can also be computed:

\[
\mathrm{CP}_{\mathrm{norm}}(e)
=
\frac{1}{|\mathcal{S}_e|}
\sum_{s\in\mathcal{S}_e} b_s(e),
\]
where
\[
\mathcal{S}_e
=
\{s\in\mathcal{S}_{\mathrm{ov}}:\exists f\in E \text{ such that } |e\cap f|\geq s\}.
\]

If \(\mathcal{S}_e=\varnothing\), we set \(\mathrm{CP}_{\mathrm{norm}}(e)=0\).

This normalized score separates persistence across meaningful overlap scales from artifacts of hyperedge size or degree.

\section{Methods}

\subsection{Hypergraph Generation via Clique Expansion}

To simulate synthetic higher-order networks with controlled structural differences, we construct hypergraphs from seed graphs by promoting maximal cliques to hyperedges using NetworkX graph generators and clique enumeration~\cite{networkx2008}. Given a seed graph \(G=(V_G,E_G)\), each maximal \(k\)-clique with \(k\geq k_{\min}\) is promoted to a hyperedge in \(\mathcal{H}=(V_G,\mathcal{E})\). We consider three widely used generative models for the seed graphs:

\begin{itemize}
    \item \textbf{Barabási--Albert (BA)} model, which captures preferential attachment and scale-free degree distributions \cite{barabasi1999emergence}.
    \item \textbf{Erdős--Rényi (ER)} model, characterized by independent edge formation with uniform edge probability \cite{erdos1959random}.
    \item \textbf{Watts--Strogatz (WS)} model, which interpolates between regular lattices and small-world networks \cite{watts1998collective}.
\end{itemize}

Because \texttt{networkx.find\_cliques} returns maximal cliques, no additional nonmaximal-hyperedge pruning is required for the clique-expanded hypergraphs. This step is therefore treated as a consistency check rather than as a substantive preprocessing operation. Connectivity is verified using the two-section graph of \(\mathcal{H}\); when the generated structure is disconnected and component filtering is enabled, the largest connected component is retained.

\medskip

Although the clique expansion method allows for a principled transformation from graph-based models to higher-order structures, it is important to acknowledge that this process can alter some of the topological characteristics of the original seed graphs. For example, the degree distribution of the resulting hypergraph may no longer strictly follow the same statistical patterns—such as power-law scaling in the Barabási–Albert model due to the overlap structure among cliques. Similarly, clustering coefficients and shortest-path characteristics may be distorted when group-based interactions are reconstructed from dyadic seeds.

Related work has shown that higher-order structure can be analyzed through simplicial closure, higher-order link prediction, and hypergraph motif profiles \cite{benson2018simplicial, lotito2022higher}. In particular, empirical higher-order studies sometimes infer group interactions from lower-order or temporal co-occurrence data, including procedures that promote observed cliques to hyperedges \cite{lotito2022higher}. These studies recognize the trade-off between analytic tractability and structural fidelity. While we do not claim that the original seed topology is perfectly preserved, the generative models provide a useful scaffolding that retains key qualitative differences---such as degree heterogeneity or rewiring locality---that manifest in the higher-order structure.

\begin{table}[H]
\centering
\begin{tabular}{lccc}
\toprule
\textbf{Parameter} & \textbf{Erdős--Rényi} & \textbf{Watts--Strogatz} & \textbf{Barabási--Albert} \\
\midrule
Seed-graph vertices \(n\) & 500 & 500 & 500 \\
ER edge probability \(p\) & 0.05 & -- & -- \\
WS neighborhood parameter & -- & 4 & -- \\
WS rewiring probability & -- & 0.10 & -- \\
BA attachment parameter \(m\) & -- & -- & 3 \\
Minimum clique size \(k_{\min}\) & 3 & 3 & 3 \\
Largest component retained & yes & yes & yes \\
Overlap scales \(\mathcal{S}_{\mathrm{ov}}\) & \(\{1,2,3\}\) & \(\{1,2,3\}\) & \(\{1,2,3\}\) \\
Number of spectral clusters \(k\) & 2 & 2 & 2 \\
Initial infected fraction & 0.10 & 0.10 & 0.10 \\
Trace-panel transmission probability \(\beta\) & 0.50 & 0.50 & 0.50 \\
Maximum timesteps \(T_{\max}\) & 50 & 50 & 50 \\
Dynamics trials shown in trace figures & 500 & 500 & 500 \\
Bifurcation scan grid & \(50\times 50\) & \(50\times 50\) & \(50\times 50\) \\
Bifurcation trials per grid point & 50 & 50 & 50 \\
Bifurcation \(\beta\) range & 0.10--0.50 & 0.10--0.50 & 0.10--0.50 \\
Bifurcation ranked-set removal-fraction range & 0.10--0.90 & 0.10--0.90 & 0.10--0.90 \\
Main reported ranked-set removal fraction & 0.50 & 0.40 & 0.10 \\
\bottomrule
\end{tabular}
\caption{Simulation and generation settings used in the reported experiments.}
\label{tab:simulation_parameters}
\end{table}

\subsection{Baseline Contagion Simulation}

We simulate a discrete-time SI-like contagion process on the hypergraph \(\mathcal{H}=(V,E)\). At time \(t=0\), an initial infected set \(I_0\subseteq V\) is selected uniformly at random with size equal to \(10\%\) of the vertex set, rounded down with at least one initial infected vertex. Once infected, a vertex remains infected.

At each timestep, every hyperedge containing at least one infected vertex becomes active. For each active hyperedge \(e\), each susceptible vertex \(v\in e\setminus I_t\) becomes infected with probability \(\beta\). Thus, infection is transmitted through active group memberships rather than through a single dyadic contact graph. If a susceptible vertex belongs to multiple active hyperedges during the same timestep, it may receive multiple independent infection opportunities.

The process stops when no new infections occur during a timestep or when \(T_{\max}=50\) timesteps have elapsed. We record the infection fraction over time and the final infected fraction across Monte Carlo trials. Unless otherwise stated, the initial infected fraction is \(0.10\) and the transmission probability is \(\beta=0.50\).

Because the process is stopped after the first timestep with no new infections, this model should be interpreted as a finite-horizon stochastic SI-like process rather than as an infinite-time SI process. In a standard repeated-attempt SI process on a connected finite structure, eventual full infection would occur with probability one for \(\beta>0\). The stopping rule used here instead captures whether the contagion continues to expand under the realized sequence of stochastic activation events.

\subsection{\(s\)-Line Graph Construction and Spectral Partitioning}

For each hypergraph \(\mathcal{H}=(V,E)\) and each \(s\in\mathcal{S}_{\mathrm{ov}}\), we construct a weighted \(s\)-line graph
\[
L_s(\mathcal{H})=(E,F_s,w_s),
\]
where
\[
\{e_i,e_j\}\in F_s
\quad\Longleftrightarrow\quad
|e_i\cap e_j|\geq s,
\]
and
\[
w_s(e_i,e_j)=|e_i\cap e_j|.
\]
Thus, the vertices of \(L_s(\mathcal{H})\) are hyperedges of the original hypergraph, and the edge weights record the magnitude of hyperedge overlap.

For each \(L_s(\mathcal{H})\), we compute the normalized Laplacian and apply spectral \(k\)-way clustering. The resulting cluster assignment \(c_s(e)\) is used to determine whether hyperedge \(e\) lies on a boundary between spectral communities. Isolated vertices in \(L_s(\mathcal{H})\) are retained as singleton vertices and assigned cluster labels by the spectral-clustering implementation; their boundary indicator is set to zero unless they have a neighbor in another cluster.

The full ranking procedure is:

\begin{enumerate}
    \item For each \(s\in\mathcal{S}_{\mathrm{ov}}\), construct \(L_s(\mathcal{H})\).
    \item Apply spectral \(k\)-way clustering to \(L_s(\mathcal{H})\).
    \item Compute \(b_s(e)\) for each hyperedge \(e\in E\).
    \item Aggregate scores using \(\mathrm{CP}(e)=\sum_{s\in\mathcal{S}_{\mathrm{ov}}}b_s(e)\).
    \item Rank hyperedges by \(\mathrm{CP}(e)\). In the reported implementation, ties are broken deterministically by hyperedge identifier. The normalized score \(\mathrm{CP}_{\mathrm{norm}}(e)\) is used as a diagnostic rather than as part of the reported ranking unless otherwise stated.
\end{enumerate}

\subsection{Contagion Suppression via Targeted Hyperedge Removal}

Let \(E_{\mathrm{rank}}\subseteq E\) denote the set of hyperedges returned by the cut-persistence ranking. In the reported implementation, hyperedges with zero cut-persistence scores are omitted from this ranked set unless otherwise specified. For a removal fraction \(r\in[0,1]\), let
\[
q_r=\lfloor r|E_{\mathrm{rank}}|\rfloor
\]
and let \(R_r\subseteq E_{\mathrm{rank}}\) denote the set of the top \(q_r\) ranked hyperedges. Targeted removal produces the edge-deleted hypergraph
\[
\mathcal{H}'_r=(V,E\setminus R_r).
\]
We then repeat the contagion simulation on \(\mathcal{H}'_r\). As a baseline, we also remove \(q_r\) hyperedges uniformly at random from the original hyperedge set \(E\).

\subsubsection{Defining Suppression Efficacy}

Let \(I(\mathcal{H})\) denote the expected final infected fraction on the original hypergraph, and let \(I(\mathcal{H}'_r)\) denote the expected final infected fraction after removing hyperedges according to removal fraction \(r\). Suppression efficacy is defined as
\[
\mathcal{S}(r)
=
\frac{I(\mathcal{H})-I(\mathcal{H}'_r)}{I(\mathcal{H})}.
\]
Thus, \(\mathcal{S}(r)>0\) indicates that the intervention reduced the final infection size, whereas \(\mathcal{S}(r)<0\) indicates that the intervention increased the final infection size.

To evaluate changes in the timing of contagion, we define the relative convergence delay
\[
\mathcal{D}_{eq}(r)
=
\frac{\tau_{eq}(\mathcal{H}'_r)-\tau_{eq}(\mathcal{H})}
{\tau_{eq}(\mathcal{H})}.
\]
Here, \(\tau_{eq}(\mathcal{H})\) and \(\tau_{eq}(\mathcal{H}'_r)\) are estimated convergence times for the original and intervened hypergraphs, respectively. A positive value of \(\mathcal{D}_{eq}(r)\) indicates slower convergence after intervention; a negative value indicates faster convergence.

In the present simulations, \(\tau_{eq}\) is estimated using the knee point of the mean infection trajectory, as determined by the Kneedle algorithm \cite{satopa2011kneedle} implemented in the \texttt{kneed} Python library \cite{kneed}. Because this knee point measures the onset of flattening rather than an absorbing-state time, we interpret \(\mathcal{D}_{eq}\) as a convergence-delay proxy rather than as an exact equilibrium time.

\section{Results}

\subsection{Summary of Intervention Outcomes Across Topologies}

The intervention outcomes show substantial topology dependence. In the Erdős--Rényi case, cut-persistence removal reduced final infection size by \(18.6\%\), compared with \(1.1\%\) for random removal. The convergence-delay metric was \(0.500\) for both removal strategies, indicating a \(50\%\) increase in the estimated convergence time rather than an acceleration.

In the Watts--Strogatz case, targeted and random removal produced very similar suppression. Targeted removal reduced final infection size by \(39.0\%\), while random removal reduced final infection size by \(40.2\%\). Neither intervention produced a measurable change in the convergence-delay metric. Thus, in this topology, the proposed cut-persistence ranking did not outperform random hyperedge deletion at the reported intervention budget.

Barabási--Albert hypergraphs exhibited the greatest robustness to hyperedge removal. Targeted removal reduced final infection size by only \(3.1\%\), whereas random removal reduced final infection size by \(7.8\%\). This suggests that, for the reported Barabási--Albert instance, hyperedge boundary structure in \(s\)-line graph partitions does not align well with dynamical influence under the contagion model used here.

Taken together, these results show that cut-persistence targeting can outperform random removal in some settings, but the advantage is not universal. The proposed score identifies structurally salient hyperedges, yet structural salience does not necessarily imply maximal contagion suppression. This motivates comparison with stronger baselines, including size-matched random removal, hyperedge-degree removal, line-graph betweenness removal, and removal curves across a common range of intervention budgets.

Values in Table~\ref{tab:suppression_metrics} are point estimates from the simulation protocol described in Table~\ref{tab:simulation_parameters}. 

\begin{table}[H]
\centering
\begin{tabular}{lcc}
\toprule
\textbf{Graph Type and Intervention} 
& \textbf{Suppression Efficacy \(\mathcal{S}\)} 
& \textbf{Convergence Delay \(\mathcal{D}_{eq}\)} \\
\midrule
Erdős--Rényi (Targeted) & 0.186 & 0.500 \\
Erdős--Rényi (Random)   & 0.011 & 0.500 \\
\addlinespace
Barabási--Albert (Targeted) & 0.031 & 0.062 \\
Barabási--Albert (Random)   & 0.078 & 0.000 \\
\addlinespace
Watts--Strogatz (Targeted) & 0.390 & 0.000 \\
Watts--Strogatz (Random)   & 0.402 & 0.000 \\
\bottomrule
\end{tabular}
\caption{Suppression efficacy \(\mathcal{S}\) and convergence delay \(\mathcal{D}_{eq}\) for each graph topology under targeted and random hyperedge-removal strategies. Targeted removal deletes high cut-persistence hyperedges from the ranked set; random removal deletes the same number of hyperedges uniformly at random from the original hyperedge set. Positive \(\mathcal{D}_{eq}\) indicates delayed convergence relative to the baseline.}
\label{tab:suppression_metrics}
\end{table}

Because the intervention budget strongly affects suppression efficacy, the single-budget comparisons in Table~\ref{tab:suppression_metrics} should be interpreted cautiously. A stronger evaluation would compare targeted and random removal across a common grid of removal fractions,
\[
r\in\{0,0.05,0.10,\dots,0.50\},
\]
and would include additional baselines such as size-matched random removal, hyperedge-size removal, hyperedge-degree removal, and line-graph-betweenness removal. We leave this removal-curve analysis for future work.

For the bifurcation scans, we vary the transmission probability \(\beta\) over the grid reported in Table~\ref{tab:simulation_parameters} and vary the ranked-set removal fraction \(r\) over the corresponding removal grid. For each pair \((\beta,r)\), hyperedges are removed according to the cut-persistence ranking, and the contagion process is simulated for the number of Monte Carlo trials reported in Table~\ref{tab:simulation_parameters}. Each heatmap summarizes the resulting final infection fraction, convergence-time proxy, and distributional statistics across trials.

\subsection{Bifurcation Scan Across Network Topologies}

\begin{figure}[H]
    \centering
    \includegraphics[width=\textwidth, height=0.6\textheight, keepaspectratio]{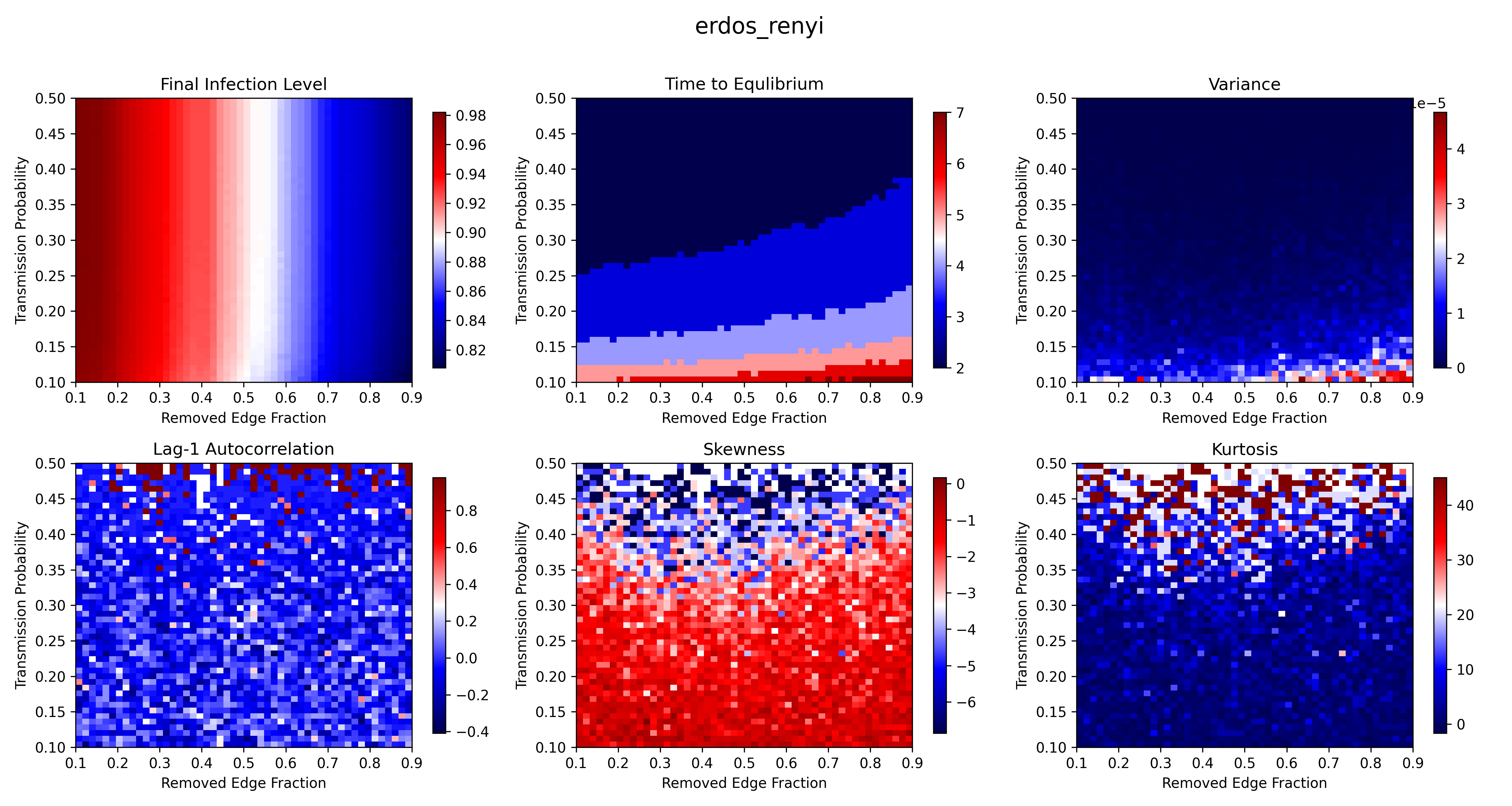}
    \caption{Bifurcation scan results for the Erdős–Rényi hypergraph under varying transmission probabilities (y-axis) and ranked-set removal fractions (x-axis). Each subplot presents a key measure of contagion or early warning signal (EWS) dynamics. \textbf{Top left:} Final infection level at equilibrium. \textbf{Top center:} Time to equilibrium, estimated via the elbow (knee) point of the mean contagion curve using the Kneedle algorithm. \textbf{Top right:} Variance in final infection levels across trials. \textbf{Bottom left:} Lag-1 autocorrelation of the infection curves, capturing short-term memory effects. \textbf{Bottom center:} Skewness, indicating the asymmetry of infection level distributions. \textbf{Bottom right:} Kurtosis, reflecting the tail weight and peak sharpness of those distributions. The scan reveals a relatively sharp phase transition in final infection levels and prolonged convergence at low transmission rates. Skewness and kurtosis vary substantially across the scan, suggesting that these quantities may be useful diagnostics for future analysis, although we do not treat them here as validated early-warning indicators. Notably, skewness and kurtosis both display an approximate U-shaped relationship along the \emph{ranked-set removal fraction axis} — that is, values are elevated at both low and high disruption levels, with a dip in the middle. This suggests that both fully intact and heavily degraded networks exhibit unstable or heterogeneous contagion outcomes.}
    \label{fig:bifurcation_erdos_renyi}
\end{figure}

\begin{figure}[H]
    \centering
    \includegraphics[width=\textwidth, height=0.6\textheight, keepaspectratio]{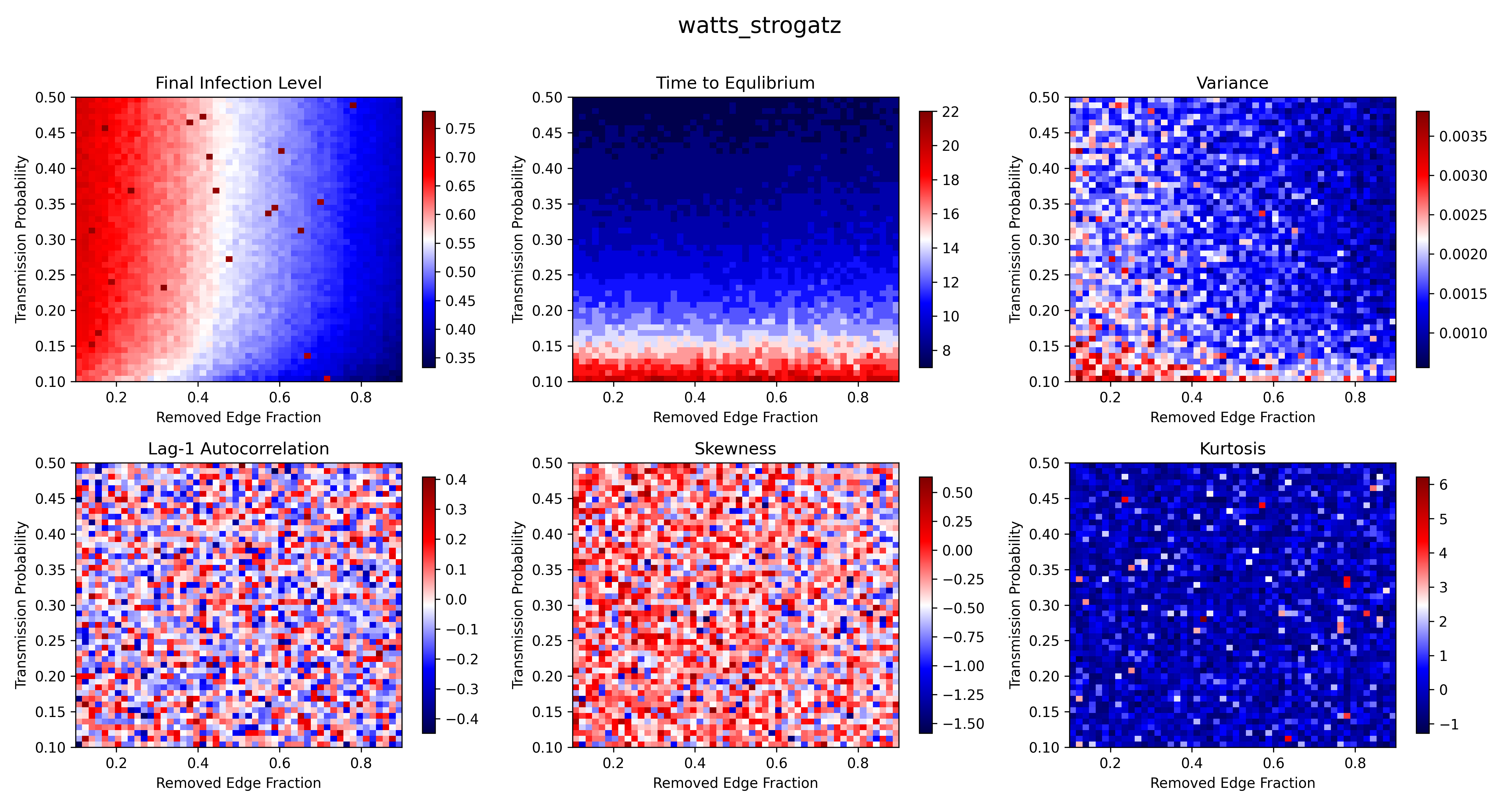}
    \caption{Bifurcation scan results for the Watts–Strogatz hypergraph under varying transmission probabilities (y-axis) and ranked-set removal fractions (x-axis). Each subplot presents a key measure of contagion or early warning signal (EWS) dynamics. \textbf{Top left:} Final infection level at equilibrium. \textbf{Top center:} Time to equilibrium, estimated using the elbow of the average contagion trajectory via the Kneedle algorithm. \textbf{Top right:} Variance in final infection levels across trials. \textbf{Bottom left:} Lag-1 autocorrelation of the infection curves, capturing short-term memory in the dynamics. \textbf{Bottom center:} Skewness, measuring asymmetry in infection distribution. \textbf{Bottom right:} Kurtosis, reflecting heavy-tailedness and peak sharpness in the distribution. The scan shows a relatively smooth phase transition in infection size with longer convergence near threshold regimes. Variance becomes elevated in parts of the low-transmission regime, suggesting parameter regions where contagion outcomes are more heterogeneous across trials.}
    \label{fig:bifurcation_watts_strogatz}
\end{figure}

\begin{figure}[H]
    \centering
    \includegraphics[width=\textwidth, height=0.6\textheight, keepaspectratio]{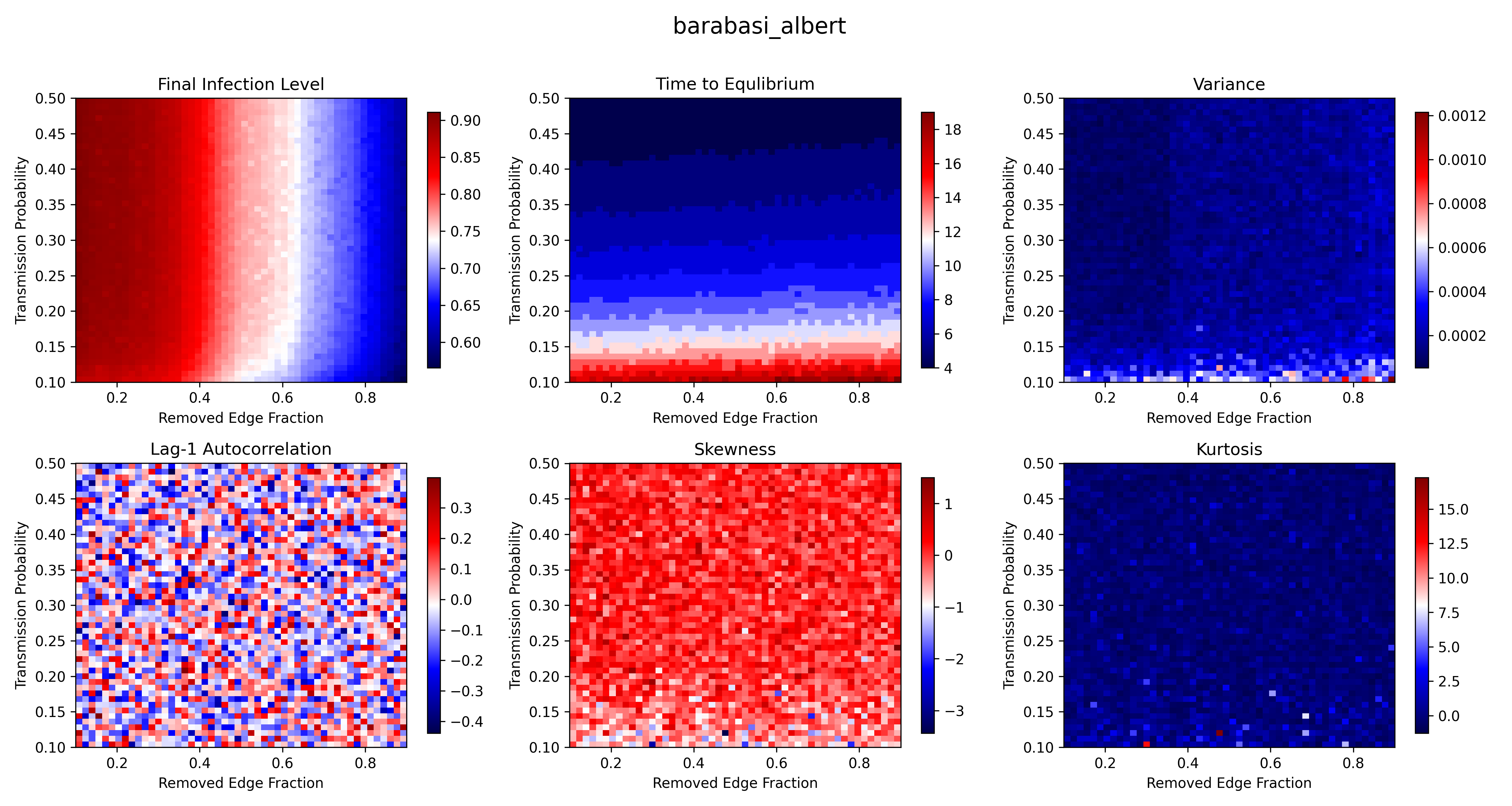}
    \caption{Bifurcation scan results for the Barabási–Albert hypergraph under varying transmission probabilities (y-axis) and ranked-set removal fractions (x-axis). Each subplot presents a key measure of contagion or early warning signal (EWS) dynamics. \textbf{Top left:} Final infection level at equilibrium. \textbf{Top center:} Time to equilibrium, estimated using the elbow of the average contagion trajectory via the Kneedle algorithm. \textbf{Top right:} Variance in final infection levels across trials. \textbf{Bottom left:} Lag-1 autocorrelation of the infection curves, capturing short-term memory effects. \textbf{Bottom center:} Skewness, measuring distribution asymmetry. \textbf{Bottom right:} Kurtosis, indicating tail weight and peakedness in infection outcomes. The scan reveals a gradual phase transition in epidemic magnitude with low early warning signal intensity and fast convergence across most of the parameter space. Within this synthetic realization, the Barabási--Albert hypergraph exhibits comparatively low sensitivity to the tested hyperedge-removal and transmission-probability ranges.}
    \label{fig:bifurcation_barabasi_albert}
\end{figure}

\subsection{Contagion Dynamics}

\begin{figure}[H]
    \centering
    \includegraphics[width=\textwidth, height=0.6\textheight, keepaspectratio]{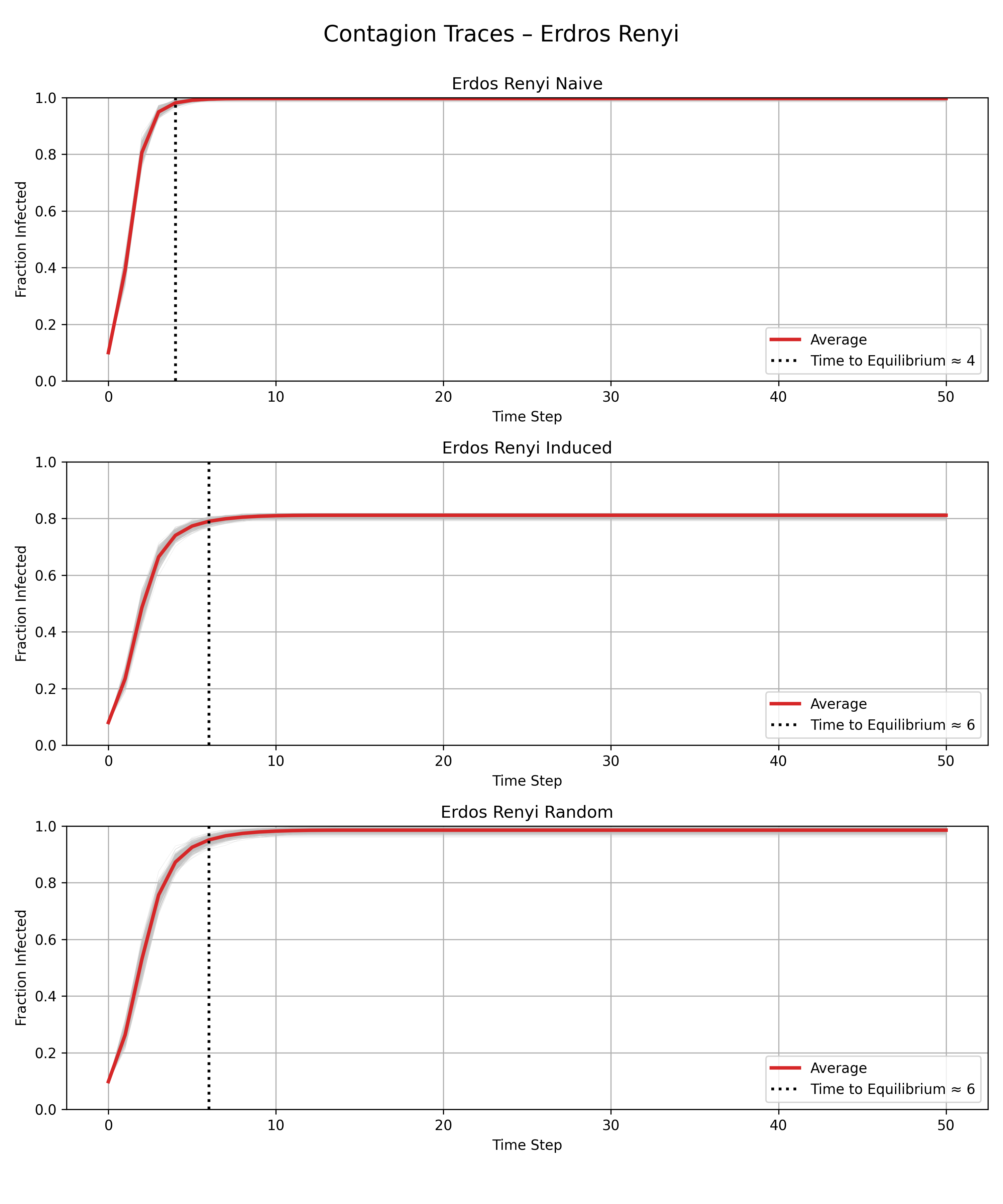}
    \caption{Contagion dynamics for Erdős–Rényi hypergraphs under three intervention regimes. Each panel shows 500 simulation traces of contagion spread over time, with the mean trajectory highlighted in red. In the intervention regimes, 50\% of the ranked high cut-persistence hyperedges were removed. \textbf{Top:} Naive contagion on the unaltered network reaches equilibrium at step 4. \textbf{Middle:} Contagion after targeted removal of high cut-persistence hyperedges identified via spectral cut-persistence scoring yields a noticeable reduction in final infection level and results in \textit{slower} convergence (step 6). \textbf{Bottom:} Random hyperedge removal results in negligible suppression and a similar slower convergence (step 6). The targeted strategy yields a suppression efficacy of \( \mathcal{S} = 0.186 \) and convergence delay \( \mathcal{D}_{eq} = 0.500 \), while random removal achieves \( \mathcal{S} = 0.011 \) and \( \mathcal{D}_{eq} = 0.500 \). These results suggest that, in Erdős–Rényi networks, structural fragmentation can reduce epidemic magnitude and slow its stabilization.}
    \label{fig:erdos_renyi_trace_dynamics}
\end{figure}

\begin{figure}[H]
    \centering
    \includegraphics[width=\textwidth, height=0.6\textheight, keepaspectratio]{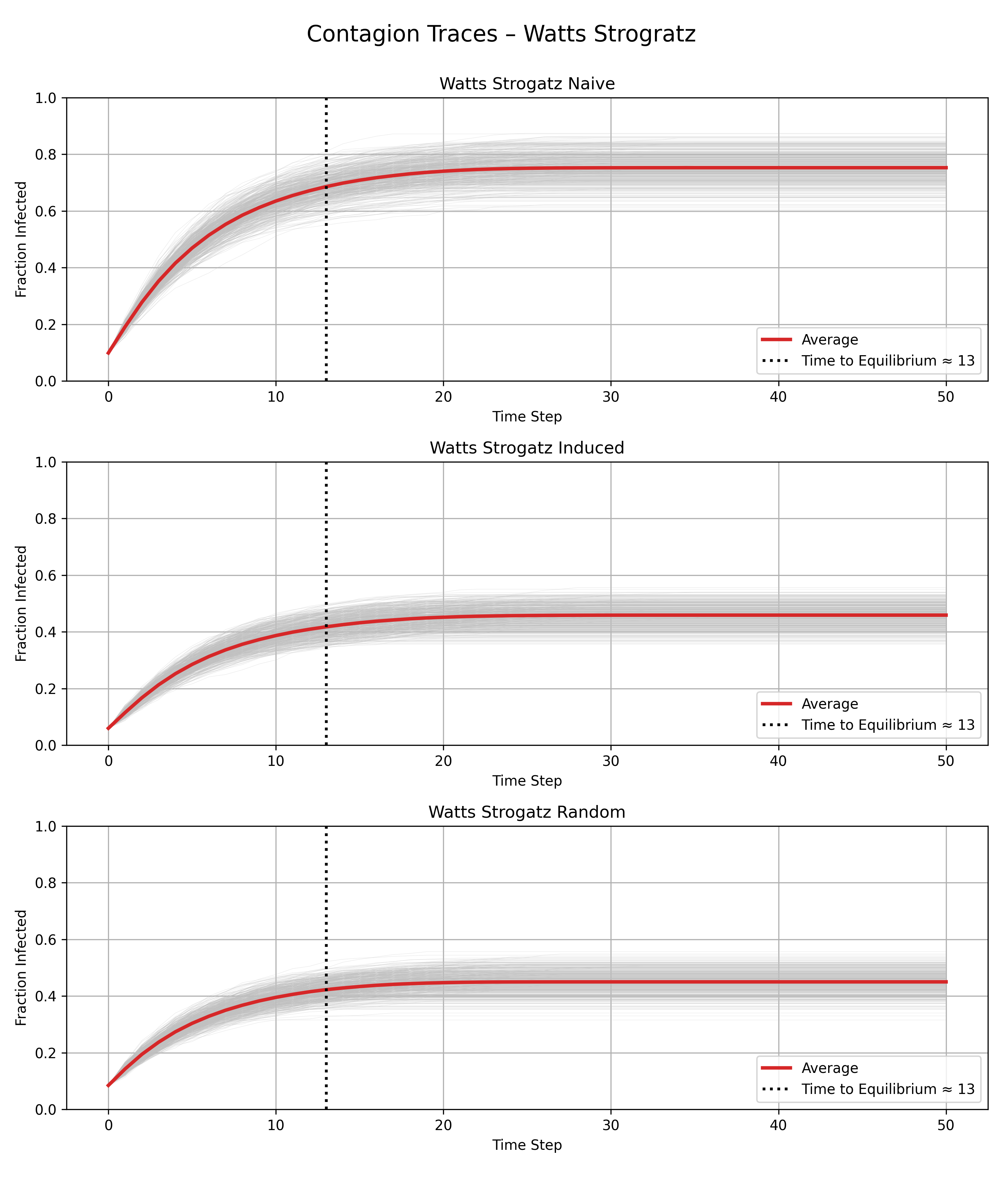}
    \caption{
    Contagion dynamics for the Watts–Strogatz hypergraph under three intervention conditions. Each panel shows 500 simulation traces of contagion spread over time, with the mean trajectory highlighted in red. In the intervention regimes, 40\% of the ranked high cut-persistence hyperedges were removed. \textbf{Top:} Naive contagion on the unaltered network, reaching approximately 73\% final infection and stabilizing near step 13. \textbf{Middle:} Contagion after targeted removal of high cut-persistence hyperedges identified via spectral cut-persistence scoring; the final infection level is suppressed by approximately 39\% (\( \mathcal{S} = 0.390 \)) with no change in convergence time (\( \mathcal{D}_{eq} = 0.000 \)). \textbf{Bottom:} Random hyperedge removal yields slightly higher suppression efficacy (\( \mathcal{S} = 0.402 \)) and also preserves the time to equilibrium (\( \mathcal{D}_{eq} = 0.000 \)). These results indicate that both targeted and random hyperedge removal reduce final infection size in this regime, with random removal performing slightly better in the reported simulations.}
    \label{fig:watts_strogatz_traces}
\end{figure}

\begin{figure}[H]
    \centering
    \includegraphics[width=\textwidth, height=0.6\textheight, keepaspectratio]{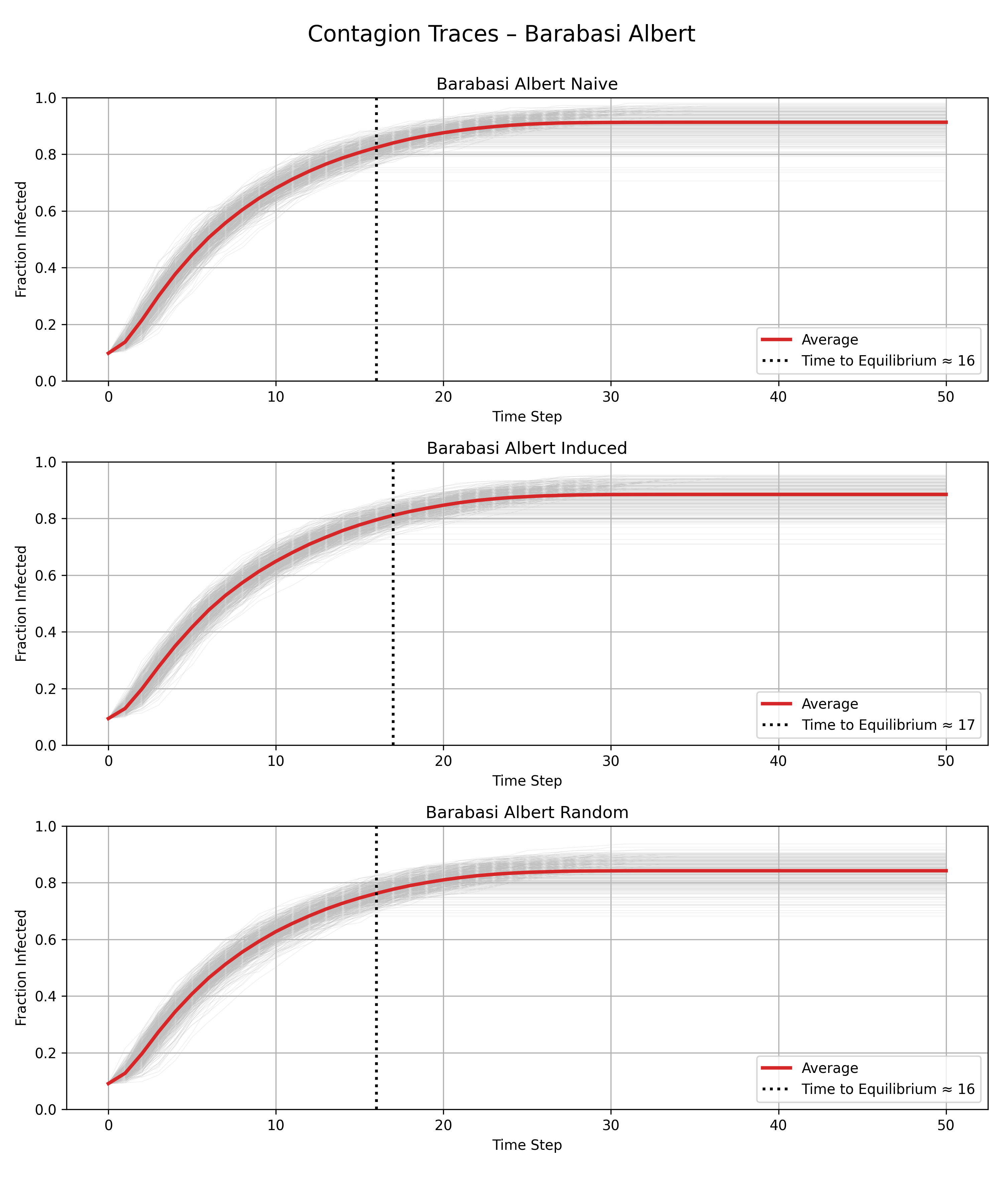}
    \caption{Contagion traces for the Barabási–Albert hypergraph under three intervention strategies. Each panel shows 500 simulation traces of contagion spread over time, with the mean trajectory highlighted in red. In the intervention regimes, 10\% of the ranked high cut-persistence hyperedges were removed. \textbf{Top:} Naive contagion process without hyperedge removal reaches equilibrium at step 16. \textbf{Middle:} Contagion after targeted removal of hyperedges with high cut-persistence scores results in a modest reduction in final infection level and a slight increase in convergence time (step 17). \textbf{Bottom:} Random hyperedge removal achieves slightly greater suppression of the final infection fraction, with no change in time to equilibrium (step 16). Suppression efficacy for the targeted strategy is \( \mathcal{S} = 0.031 \) and \( \mathcal{D}_{eq} = 0.062 \), while the random strategy yields \( \mathcal{S} = 0.078 \) and \( \mathcal{D}_{eq} = 0.000 \). These results suggest that this synthetic Barabási--Albert realization is comparatively insensitive to the tested targeted and random hyperedge-removal interventions.}
    \label{fig:barabasi_albert_traces}
\end{figure}

\section{Discussion}

This study proposes cut persistence as a spectral score for ranking hyperedges according to their boundary role in \(s\)-line graph partitions. The simulations show that removing high-scoring hyperedges can suppress contagion in some synthetic topologies, but the advantage over random removal is not universal. Thus, the results should be interpreted as a proof of concept for hyperedge-overlap-based intervention rather than as evidence of a generally optimal dismantling strategy.

In the Erdős--Rényi case, cut-persistence targeting produced a larger reduction in final infection size than random removal. This suggests that, in relatively homogeneous overlap structures, spectral boundary hyperedges may correspond to dynamically important transmission pathways. However, both targeted and random removal produced the same reported convergence-delay value, so the temporal effect should not be attributed uniquely to the spectral targeting procedure.

In the Watts--Strogatz case, targeted and random removal produced nearly identical suppression, with random removal slightly higher in the reported simulations. This indicates that the observed suppression may be driven by the overall removal budget rather than by the specific cut-persistence ranking. Additional baselines, including size-matched random removal and line-graph-degree removal, are needed before concluding that spectral targeting adds value in small-world-like hypergraphs.

In the Barabási--Albert case, random removal outperformed targeted removal. This is consistent with the possibility that the contagion dynamics in hub-dominated structures are controlled more by node-level centrality or incidence concentration than by boundary structure in hyperedge-overlap partitions. An edge-centric intervention may therefore miss the dominant routes of transmission in scale-free or highly heterogeneous systems.

Overall, these findings show that hyperedge-overlap structure is relevant to intervention design, but also that structural salience and dynamical influence are distinct quantities. Future work should evaluate the proposed score across ensembles of independently generated hypergraphs, report uncertainty intervals, compare against stronger baselines, and test contagion rules with explicitly higher-order reinforcement or threshold effects.

\section{Conclusions and Future Work}

This work introduces a framework for ranking hyperedges using spectral \(k\)-way partitioning of \(s\)-line graphs. The resulting cut-persistence score identifies hyperedges that repeatedly lie on boundaries between spectral communities across multiple overlap scales.

The empirical results show that the impact of hyperedge removal is topology-dependent. In the reported Erdős--Rényi case, cut-persistence targeting outperformed random removal in reducing final infection size. In the Watts--Strogatz and Barabási--Albert cases, however, random removal was comparable to or better than targeted removal. These findings indicate that cut persistence can identify structurally salient hyperedges, but that such hyperedges are not necessarily the most dynamically influential under all contagion models or network topologies.

Several extensions are necessary before the method can be treated as a general intervention strategy. First, future work should evaluate removal curves over a common range of removal fractions rather than selecting a single intervention budget for each topology. Second, the method should be compared with stronger baselines, including hyperedge-size removal, hyperedge-degree removal, line-graph betweenness removal, and size-matched random removal. Third, experiments should be repeated over ensembles of independently generated hypergraphs, with uncertainty intervals reported across both graph realizations and contagion trials.

Fourth, future studies should test genuinely higher-order contagion mechanisms. The present model is an SI-like stochastic contagion process with group-activated hyperedge transmission, but it does not include threshold, reinforcement, or nonlinear group-contagion effects. Threshold, reinforcement, nonlinear group-transmission, SIR, or SIS dynamics may interact with hypergraph topology in qualitatively different ways. Finally, empirical hypergraph datasets should be used to evaluate whether cut-persistence rankings identify meaningful intervention targets in real systems.

In sum, the present study provides a useful starting point for spectral hypergraph intervention, but its current evidence supports a more limited conclusion: cut-persistence targeting is a promising structural heuristic whose dynamical effectiveness depends on topology, contagion mechanism, and comparison baseline.

\paragraph{Reproducibility.}
Reproducibility code is available at
\url{https://github.com/DemetersSon83/Hypergraph-Edge-Disintegration}.
The repository contains code for hypergraph generation, \(s\)-line graph construction, spectral clustering, cut-persistence ranking, SI-like contagion simulation, and plotting. The implementation uses Python with NumPy, SciPy, pandas, NetworkX, matplotlib, scikit-learn, statsmodels, kneed, and PyYAML.

\section{Acknowledgments}
The authors would like to thank Dr. Emma Zajdela for helpful discussions.

\bibliographystyle{plain}
\bibliography{bibliography}

\end{document}